\documentclass[12pt]{article}

\usepackage{amsmath, amssymb}
\usepackage{graphicx}
\usepackage{hyperref}
\usepackage{placeins}
\usepackage{natbib}
\usepackage{geometry}
\usepackage{float}
\usepackage[font={small,it}]{caption}
\usepackage{setspace}
\usepackage{newtxtext,newtxmath} 
\usepackage{indentfirst}   
\title{\textbf{Bitcoin Forecasting with Classical Time Series Models on Prices and Volatility}}
\author{
\begin{tabular}{c}
Anmar Kareem \quad | \quad Dr.~Alexander Aue \\
Department of Statistics and Data Science \\
University of California, Davis
\end{tabular}
}
\date{September 11, 2025}

\begin{document}

\maketitle

\begin{abstract}
\begin{spacing}{1.75}
\noindent
This paper evaluates the performance of classical time series models in forecasting Bitcoin prices, focusing on ARIMA, SARIMA, GARCH, and EGARCH. Daily price data from 2010 to 2020 were used, with each model trained on the first 90\% and tested on the final 10\%. Forecast accuracy was assessed using MAE, RMSE, AIC, and BIC. The results show that ARIMA provided the strongest forecasts for short-run log-price dynamics, while EGARCH offered the best fit for volatility by capturing asymmetry in responses to shocks. These findings suggest that although Bitcoin’s extreme volatility remains difficult to fully capture, classical time series models can still provide valuable short-run forecasts. The study contributes to understanding cryptocurrency forecasting and sets the stage for future work that incorporates machine learning methods and broader macroeconomic or blockchain variables.  

\vspace{1em}
\end{spacing}
\end{abstract}

\newpage
\section{Introduction}
Bitcoin has become the most widely known cryptocurrency and one of the most volatile financial assets in the world. Over the past decade, its price has grown from just a few dollars to thousands, with large swings that make it both attractive and risky for investors (\hyperref[ref:pichl2017]{Pichl and Kaizoji, 2017}). This volatility has drawn attention not only from traders but also from researchers who see Bitcoin as an important case for testing forecasting methods, making it a growing focus in statistics and data science (\hyperref[ref:yenidogan2018]{Yenidoğan et al., 2018}). Understanding how to predict its price is important because sharp movements in value affect not only individual investors but also the broader financial system, where cryptocurrencies are playing a growing role.

Bitcoin data often displays two important challenges for forecasting: stationarity and seasonality. Stationarity means that the statistical properties of a process, such as its mean and variance, remain constant over time, while seasonality refers to repeating cycles that can occur daily, weekly, or at other intervals. To address these issues, researchers frequently apply classical time series techniques such as the Autoregressive Integrated Moving Average (ARIMA) model, which is designed to capture autocorrelation, differences, and seasonal patterns in the data (\hyperref[ref:yenidogan2018]{Yenidoğan et al., 2018}). In addition to these concerns, Bitcoin prices also exhibit volatility clustering, where periods of large price changes are followed by more large changes. This feature motivates the use of models such as the Generalized Autoregressive Conditional Heteroskedasticity (GARCH) family, which focus on capturing and forecasting changing levels of volatility (\hyperref[ref:pichl2017]{Pichl and Kaizoji, 2017}).

This study builds on these insights by applying a set of classical time series models to Bitcoin price data. Specifically, we employ the ARIMA, the Seasonal ARIMA (SARIMA), the GARCH, and the Exponential GARCH (EGARCH) models, training each model on the first 90\% of the dataset and forecasting the remaining 10\% to allow for an out-of-sample assessment of predictive accuracy. The performance of these approaches is evaluated using four widely applied forecasting metrics: the Mean Absolute Error (MAE), which measures the average magnitude of prediction errors in absolute terms; the Root Mean Squared Error (RMSE), which emphasizes larger errors by squaring them before averaging; the Akaike Information Criterion (AIC); and the Bayesian Information Criterion (BIC), both of which provide measures of model fit that penalize complexity. Despite Bitcoin’s complexity and volatility, classical time series models can still provide meaningful short-run forecasts when carefully applied, and by comparing ARIMA, SARIMA, GARCH, and EGARCH on a consistent dataset and evaluation framework, the study highlights both the strengths and limitations of traditional statistical methods for cryptocurrency forecasting while offering a foundation for future research that may incorporate more advanced machine learning techniques.

\section{Literature Review}
Research on Bitcoin forecasting has expanded as the asset’s volatility continues to attract attention from both finance and data science. 
This study is most directly related to the work of \hyperref[ref:mudassir2020]{Mudassir et al. (2020)}, who applied a range of machine learning models such as artificial neural networks (ANN), long short-term memory (LSTM), and support vector machines (SVM) to Bitcoin data. 
Their analysis evaluated performance across several horizons, showing that machine learning methods can sometimes outperform traditional approaches reported in the broader literature. 
Building on this, my research applies purely classical time series techniques such as ARIMA, SARIMA, GARCH, and EGARCH, while differing in design by training on the first 90\% of the dataset and forecasting the final 10\%. 
This approach allows for a focused evaluation of how each method performs at the end of the observed period.

\section{Methodology}
\subsection{Data}

The dataset used in this study is publicly available on GitHub 
(\href{https://github.com/heliphix/btc_data/blob/main/btc_data.csv}{btc\_data.csv}). 
It covers daily Bitcoin activity from July 17, 2010 to February 2, 2020, with 3,488 rows of observations. 
The data includes a wide range of blockchain and market features, organized into columns that track Bitcoin price, mining difficulty, hashrate, daily transactions, transaction fees, active addresses, and block size. 
While the dataset contains hundreds of technical indicators derived from moving averages, variance, momentum, and similar calculations, only the raw prefix metrics were used in this analysis. 
To focus on meaningful price movements, the dataset was also filtered to include only Bitcoin values of \$100 or higher. This filter was applied to remove the earliest period when Bitcoin traded at extremely low values, 
which could distort the analysis and reduce the reliability of the forecasting results.

\subsection{Preprocessing}
\subsubsection{Log Transformation}

\begin{figure}[H]
    \centering
    \includegraphics[width=\textwidth]{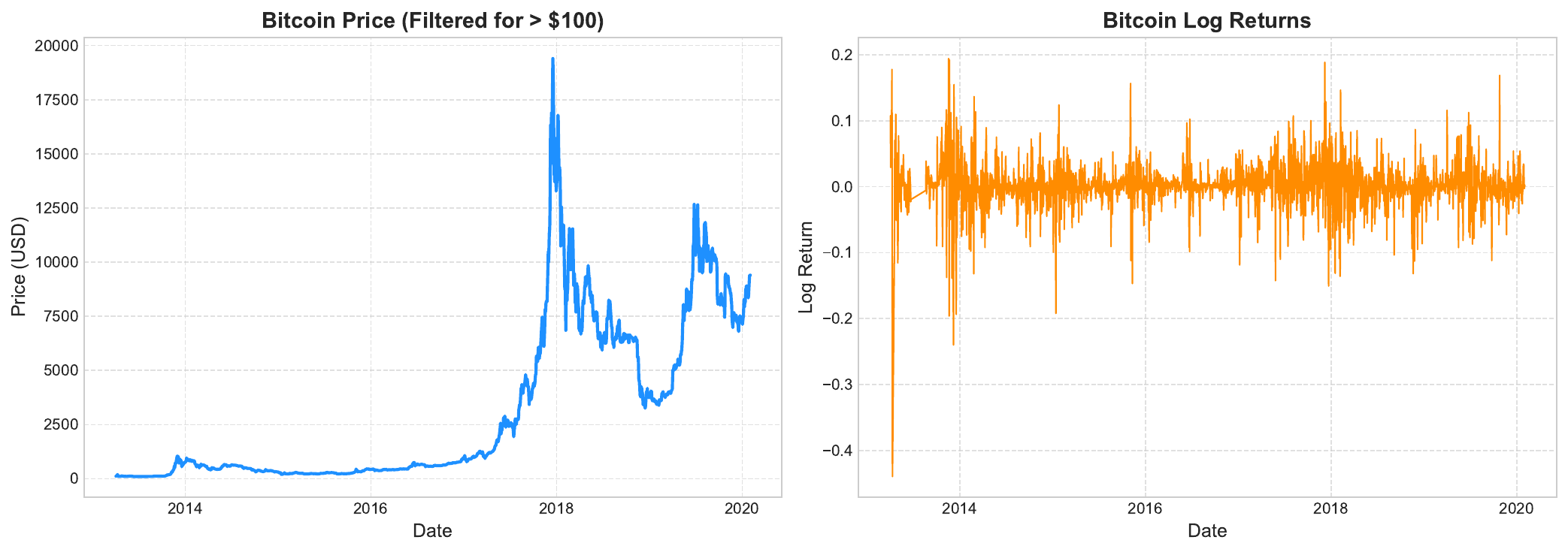}
    \caption{Bitcoin price series (left) and log returns (right).}
    \label{fig:btc_price_logreturns}
\end{figure}
The raw Bitcoin price series in Figure~\ref{fig:btc_price_logreturns} shows a steady upward trend over time, with clear explosive growth in later years. The increase is not smooth, however, as there are sharp jumps and sudden drops that reflect the volatile nature of the asset. From simple observation, the series appears highly unstable, with long upward trends and sudden corrections, making it challenging to analyze in its raw form.
\begin{equation}
\Delta \ln(P_t) = \ln(P_t) - \ln(P_{t-1})
\end{equation}

The log return series is defined in equation (1). In Figure~\ref{fig:btc_price_logreturns}, the log return series provides a clearer view of the underlying behavior. Unlike the raw prices, the log returns fluctuate around a constant mean and highlight specific periods of high volatility. In particular, large swings can be seen in 2013–2014 and again in 2017–2018, which align with the major Bitcoin bull run in 2017 and the subsequent crash in 2018. On average, the log return over the full dataset is approximately 0.001837, which suggests small positive daily returns but with substantial variation from day to day. This transformation produces a series more suitable for forecasting because it better reflects the short-term changes in Bitcoin’s value.

\begin{figure}[htbp]
    \centering
    \includegraphics[width=\textwidth]{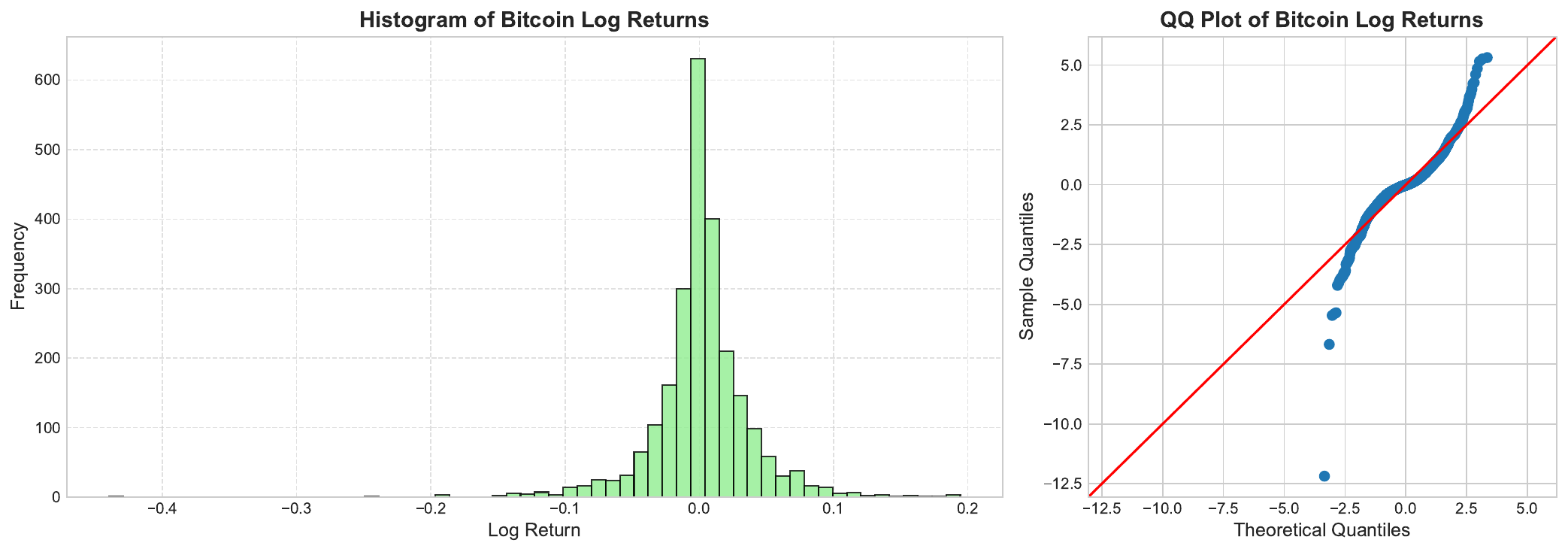}
    \caption{\textit{Histogram of Bitcoin log returns (left) and QQ plot (right).}}
    \label{fig:btc_logreturns_histqq}
\end{figure}
The histogram of Bitcoin log returns in Figure~\ref{fig:btc_logreturns_histqq} shows that most values are tightly concentrated around zero, but the distribution has noticeable outliers on both tails. This suggests that while small daily changes are common, extreme positive or negative returns occur more frequently than would be expected under a normal distribution.

The QQ plot in Figure~\ref{fig:btc_logreturns_histqq} further confirms this pattern. If the returns were normally distributed, the points would fall along the straight reference line. Instead, the observed quantiles diverge strongly in both tails, especially in the lower tail, indicating heavy-tailed behavior. In practical terms, this means Bitcoin returns exhibit a higher probability of extreme events compared to a normal distribution, which is consistent with the volatility characteristics noted in prior studies.

Together, these results support the use of log returns rather than raw prices as the input for the time series models applied in the next sections.

\subsubsection{Stationarity and Heteroskedasticity Tests}
\begin{figure}[H]
    \centering
    \includegraphics[width=\textwidth]{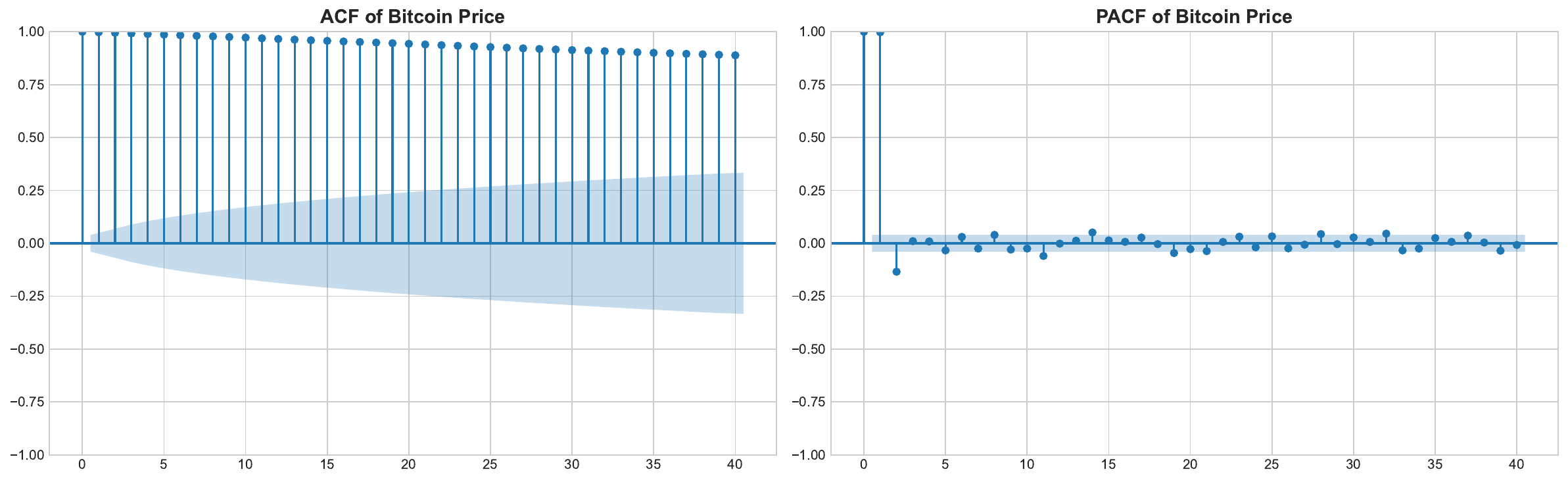}
    \caption{\textit{ACF and PACF plots for the Bitcoin price series.}}
    \label{fig:acf_pacf_price}
\end{figure}

The ACF and PACF plots in Figure~\ref{fig:acf_pacf_price} for the raw Bitcoin price show strong, persistent positive correlations that decay very slowly. This is typical of a random-walk type process, where past values strongly influence future ones and shocks do not fade quickly. Such behavior suggests the series is not stationary in its raw form. To make the data more suitable for modeling, we apply differencing, which means subtracting the previous value from the current value to remove trends and stabilize the mean.

\begin{figure}[htbp]
    \centering
    \includegraphics[width=\textwidth]{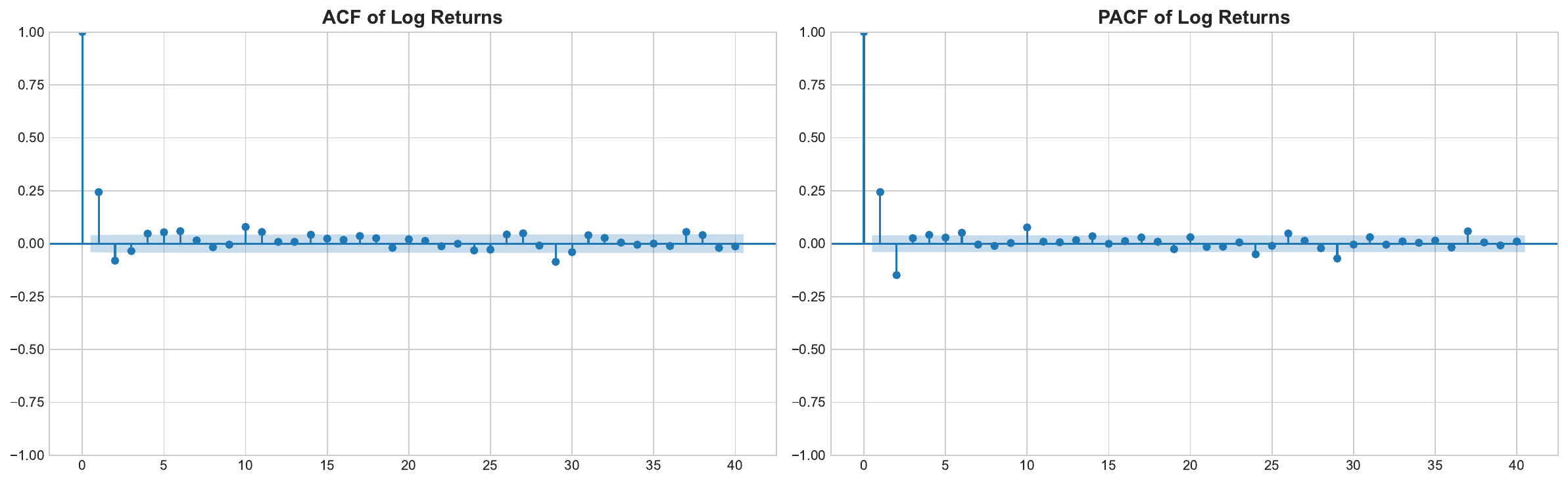}
    \caption{\textit{ACF and PACF plots for Bitcoin log returns.}}
    \label{fig:acf_pacf_logreturns}
\end{figure}

The ACF and PACF plots in Figure~\ref{fig:acf_pacf_logreturns} for Bitcoin log returns show values much closer to zero beyond the first lags, with only small, scattered spikes. This indicates that the log return series behaves more like white noise, where past values provide little information about the future. While the mean dynamics appear weak, there are still signs of volatility clustering, meaning periods of large and small changes tend to group together. This motivates the use of models that explicitly capture and forecast volatility.

\begin{table}[htbp]
    \centering
    \begin{tabular}{l c}
        \hline
        \textbf{Test} & \textbf{p-value} \\
        \hline
        Augmented Dickey--Fuller (ADF) & $3.86 \times 10^{-25}$ \\
        Engle's ARCH & $1.60 \times 10^{-84}$ \\
        \hline
    \end{tabular}
    \caption{\textit{Summary of stationarity and heteroskedasticity test results for Bitcoin log returns.}}
    \label{tab:tests}
\end{table}

In Table~\ref{tab:tests}, the ADF test produced a very low p-value, which allows us to reject the null hypothesis of a unit root. This indicates that the log return series is stationary, meaning it fluctuates around a constant mean without a persistent trend. Stationarity is a key requirement for autoregressive models, and this result justifies the use of ARIMA-type models to capture the autocorrelation structure in the data.

The Engle’s ARCH test in Table~\ref{tab:tests} also reported a highly significant p-value, providing strong evidence of conditional heteroskedasticity in the log returns. In other words, periods of high volatility tend to cluster together rather than appearing randomly. This behavior cannot be modeled adequately by ARIMA alone, which assumes constant variance in the residuals. Instead, it motivates the use of GARCH-type models, which are designed to explicitly capture and forecast volatility clustering in financial time series.

\subsubsection{Autocorrelation and Cointegration Diagnostics}

During the exploratory phase of the analysis, three sets of regression models were estimated using the raw price, the logarithm of price, and the log-differenced price of Bitcoin as the dependent variable. Each specification included blockchain-related predictors such as transaction fees, hashrate, and mining difficulty. The strongest regression from each category was selected based on adjusted $R^2$, and the resulting models are presented in the tables below. While these regressions provide some explanatory power, a major issue is that their residuals exhibit autocorrelation, which violates the assumptions of ordinary least squares and undermines stationarity. The next step is therefore to examine how this problem can be addressed and why more advanced time series models are needed.

\begin{table}[htbp]
    \centering
    \begin{tabular}{l c c}
        \hline
        \textbf{Regression} & \textbf{DW Before} & \textbf{DW After} \\
        \hline
        $price_{USD} \sim difficulty + transactionfees_{USD}$ & 0.04 & 1.50 \\
        $\log(price_{USD}) \sim \log(hashrate) + \log(difficulty)$ & & \\
        $\quad + \log(transactionfees_{USD})$ & 0.09 & 0.99 \\
        $\Delta \log(price_{USD}) \sim \Delta \log(transactionfees_{USD})$ & 1.56 & 1.90 \\
        \hline
    \end{tabular}
    \caption{\textit{Durbin--Watson statistics before and after Cochrane--Orcutt correction for selected regression models.}}
    \label{tab:dw_results}
\end{table}

The Cochrane–Orcutt procedure corrects for autocorrelation in Table~\ref{tab:dw_results} by transforming the regression model so that the error terms no longer depend strongly on their past values. In practice, this adjustment moves the Durbin–Watson statistic much closer to the desired value of 2, with the best model reaching 1.90, which indicates that residual autocorrelation has been largely removed. However, this approach only patches the problem rather than fully modeling the time-dependent structure of the data, which gives another reason to rely on classical time series models that can capture autocorrelation and volatility more systematically.

Cointegration tests were also conducted between Bitcoin prices and key blockchain variables such as hashrate, mining difficulty, and transaction fees. 
The Johansen procedure suggested the presence of at least one cointegrating relationship, implying that these variables share some degree of long-run equilibrium. 
However, since the focus of this study is on forecasting short-run dynamics using univariate time series methods, cointegration plays only a limited role in the analysis. 
Nevertheless, the results provide supporting evidence that Bitcoin prices and network fundamentals are linked over the long run, even though short-run modeling is better addressed by classical time series approaches.

\subsection{Models}
\subsubsection{ARIMA and SARIMA}

\begin{table}[htbp]
    \centering
    \begin{tabular}{l c c c}
        \hline
        \textbf{Parameter} & \textbf{Estimate} & \textbf{Std. Error} & \textbf{p-value} \\
        \hline
        AR(1)  & -0.1203 & 0.041 & 0.004 \\
        MA(1)  &  0.4103 & 0.040 & 0.000 \\
        \hline
    \end{tabular}
    \caption{\textit{Coefficient estimates for ARIMA(1,1,1).}}
    \label{tab:arima111_coef}
\end{table}

In this study, the ARIMA(1,1,1) model was chosen as a baseline because it is a simple yet flexible specification that combines one autoregressive term, one differencing step to ensure stationarity, and one moving average term. This structure is widely used in financial time series as it captures both short-run dependence and shock effects in a parsimonious way. In Table~\ref{tab:arima111_coef}, the AR(1) coefficient of –0.1203 suggests that Bitcoin’s daily price changes show a slight tendency to reverse direction, meaning an increase on one day is often followed by a small decrease the next, and vice versa. At the same time, the MA(1) coefficient of 0.4103 indicates that unexpected shocks in price tend to carry over into the following day in the same direction. Taken together, these results show that while Bitcoin prices sometimes correct themselves after a change, short-term shocks have a lingering effect that continues to influence movements from one day to the next.

\begin{table}[htbp]
    \centering
    \begin{tabular}{l c c c}
        \hline
        \textbf{Parameter} & \textbf{Estimate} & \textbf{Std. Error} & \textbf{p-value} \\
        \hline
        AR(1)  & -0.0881 & 0.043 & 0.041 \\
        MA(1)  &  0.3811 & 0.043 & 0.000 \\
        Seasonal AR(7) & -0.0090 & 0.014 & 0.521 \\
        Seasonal MA(7) & -0.9949 & 0.007 & 0.000 \\
        \hline
    \end{tabular}
    \caption{\textit{Coefficient estimates for SARIMA(1,1,1)(1,1,1,7).}}
    \label{tab:sarima1111117_coef}
\end{table}

The AR(1) coefficient of –0.0881 in Table~\ref{tab:sarima1111117_coef} indicates that Bitcoin’s daily price changes have a slight tendency to reverse direction from one day to the next, although the effect is weak. The MA(1) coefficient of 0.3811 shows that short-term shocks continue to influence the following day in the same direction, similar to the ARIMA results. The seasonal AR(7) term is very close to zero and not statistically significant, suggesting that weekly price changes do not strongly depend on the previous week’s values. However, the seasonal MA(7) coefficient of –0.9949 is highly significant, implying that unexpected shocks in one week are almost completely offset by opposite shocks in the following week. Taken together, these results suggest that while day-to-day dynamics are similar to the ARIMA model, the main seasonal pattern is not in the prices themselves but in how shocks are corrected over weekly cycles.

\subsubsection{GARCH and EGARCH}

\begin{table}[htbp]
    \centering
    \begin{tabular}{l c}
        \hline
        \textbf{Parameter} & \textbf{Estimate} \\
        \hline
        $\omega$ & $2.7147 \times 10^{-5}$ \\
        $\alpha_1$ & $0.2000$ \\
        $\beta_1$ & $0.7800$ \\
        $\alpha_1 + \beta_1$ & $0.9800$ \\
        \hline
    \end{tabular}
    \caption{\textit{Estimated parameters of the GARCH(1,1) volatility equation.}}
    \label{tab:garch_params}
\end{table}

In Table~\ref{tab:garch_params}, the parameter $\omega = 2.7147 \times 10^{-5}$ represents the long-run average volatility of Bitcoin returns. 
Although the value is small, it serves as the baseline level around which volatility fluctuates. 
The coefficient $\alpha_1 = 0.2000$ shows that short-term shocks, such as sudden price jumps or drops, have an immediate and meaningful effect on volatility. 
The coefficient $\beta_1 = 0.7800$ indicates strong persistence, meaning periods of high volatility tend to be followed by more high volatility, and calm periods by more calm periods. 
Together, $\alpha_1 + \beta_1 = 0.98$ is very close to 1, which confirms that Bitcoin returns exhibit strong volatility clustering---large price swings are not isolated events but tend to come in waves, a common feature in financial markets.

\begin{equation}
\log(\sigma_t^2) = \omega + \beta_1 \log(\sigma_{t-1}^2) 
+ \alpha_1 \left( \frac{|\varepsilon_{t-1}|}{\sigma_{t-1}} - \sqrt{\frac{2}{\pi}} \right) 
+ \gamma_1 \frac{\varepsilon_{t-1}}{\sigma_{t-1}}
\end{equation}

\begin{table}[htbp]
    \centering
    \begin{tabular}{l c}
        \hline
        \textbf{Parameter} & \textbf{Estimate} \\
        \hline
        $\omega$ & $-0.6317$ \\
        $\alpha_1$ & $0.4407$ \\
        $\beta_1$ & $0.9054$ \\
        $\gamma_1$ & \text{(negative, captures asymmetry)} \\
        \hline
    \end{tabular}
    \caption{\textit{Estimated parameters of the EGARCH(1,1) volatility equation.}}
    \label{tab:egarch_params}
\end{table}

Equation (2) specifies the EGARCH(1,1) model, which allows volatility to respond asymmetrically to positive and negative shocks. 
The parameter estimates are reported in Table~\ref{tab:egarch_params}. 
The intercept $\omega = -0.6317$ is negative, which is typical in EGARCH since the equation is expressed in logarithmic form. 
The coefficient $\alpha_1 = 0.4407$ indicates that short-run shocks have a substantial impact on volatility, capturing the immediate effect of sudden price movements. 
The persistence parameter $\beta_1 = 0.9054$ is very close to one, showing that volatility is highly persistent and that periods of high or low volatility tend to last over time. 
Finally, the asymmetry parameter $\gamma_1$ is negative, which means that negative shocks or bad news increase volatility more than positive shocks of the same size. 
This asymmetric response is a realistic feature of financial data, especially in cryptocurrency markets where downturns are often accompanied by heightened uncertainty and larger swings in price.

\subsection{Evaluation Metrics}
\begin{table}[H]
    \centering
    \begin{tabular}{l c c}
        \hline
        \textbf{Model} & \textbf{MAE} & \textbf{RMSE} \\
        \hline
        ARIMA(1,1,1) & 0.1366 & 0.1634 \\
        SARIMA(1,1,1)(1,1,1,7) & 0.2635 & 0.3228 \\
        \hline
    \end{tabular}
    \caption{\textit{Forecast accuracy metrics for ARIMA and SARIMA.}}
    \label{tab:arima_sarima_metrics}
\end{table}

Between the two models in Table~\ref{tab:arima_sarima_metrics}, ARIMA(1,1,1) performed better with lower MAE and RMSE values, showing that a simpler specification captured the short-run dynamics of Bitcoin more effectively than SARIMA. The SARIMA process added a seasonal component to account for weekly cycles, and the weekly period was found to be the best choice compared to biweekly or monthly alternatives. Longer seasonal periods of 30 days or more were not only less accurate but also computationally intensive, making the weekly model the most practical among the seasonal options. 

\begin{table}[htbp]
    \centering
    \begin{tabular}{l c c}
        \hline
        \textbf{Model} & \textbf{AIC} & \textbf{BIC} \\
        \hline
        GARCH(1,1)  & -9347.49 & -9324.74 \\
        EGARCH(1,1) & -9368.50 & -9345.74 \\
        \hline
    \end{tabular}
    \caption{\textit{Model selection criteria for GARCH and EGARCH.}}
    \label{tab:garch_egarch_aicbic}
\end{table}

Table~\ref{tab:garch_egarch_aicbic} shows that the EGARCH(1,1) model achieves lower AIC and BIC values compared to GARCH(1,1), indicating a better overall fit. This suggests that allowing for asymmetric effects in volatility provides a more accurate description of Bitcoin’s return dynamics. In particular, EGARCH captures the fact that negative shocks tend to increase volatility more than positive shocks, a pattern consistent with financial and cryptocurrency markets. By contrast, the inclusion of seasonal terms in SARIMA may have overfit noise, reducing forecast accuracy.

\section{Results}
\begin{figure}[H]
    \centering
    \includegraphics[width=\textwidth]{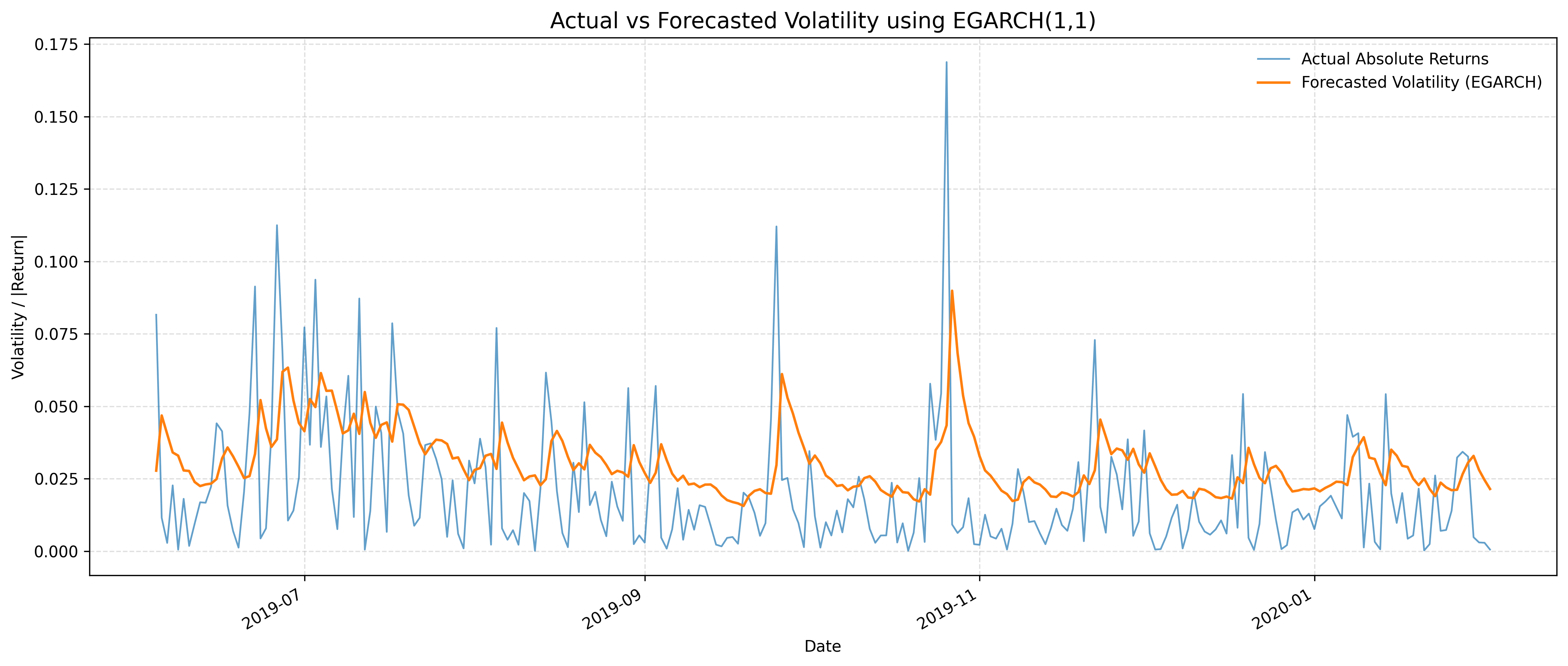}
    \caption{\textit{Actual vs forecasted volatility using EGARCH(1,1).}}
    \label{fig:volatility}
\end{figure}

Figure~\ref{fig:volatility} presents the one-step-ahead volatility forecasts from the EGARCH(1,1) model compared to the actual absolute returns. At the start of the test period, the forecasts appear somewhat rough, but the model quickly adjusts and begins to track the overall volatility dynamics more closely. The forecasted volatility rises during turbulent periods and falls during calmer ones, which is consistent with expectations from an EGARCH specification. While the model does not fully capture the largest spikes in returns, this is normal since GARCH-type models are designed to smooth volatility rather than replicate every extreme jump. Overall, the forecasts reflect the clustering and persistence of volatility in Bitcoin returns, providing a reasonable fit for the purpose of this analysis.

\begin{figure}[H]
    \centering
    \includegraphics[width=\textwidth]{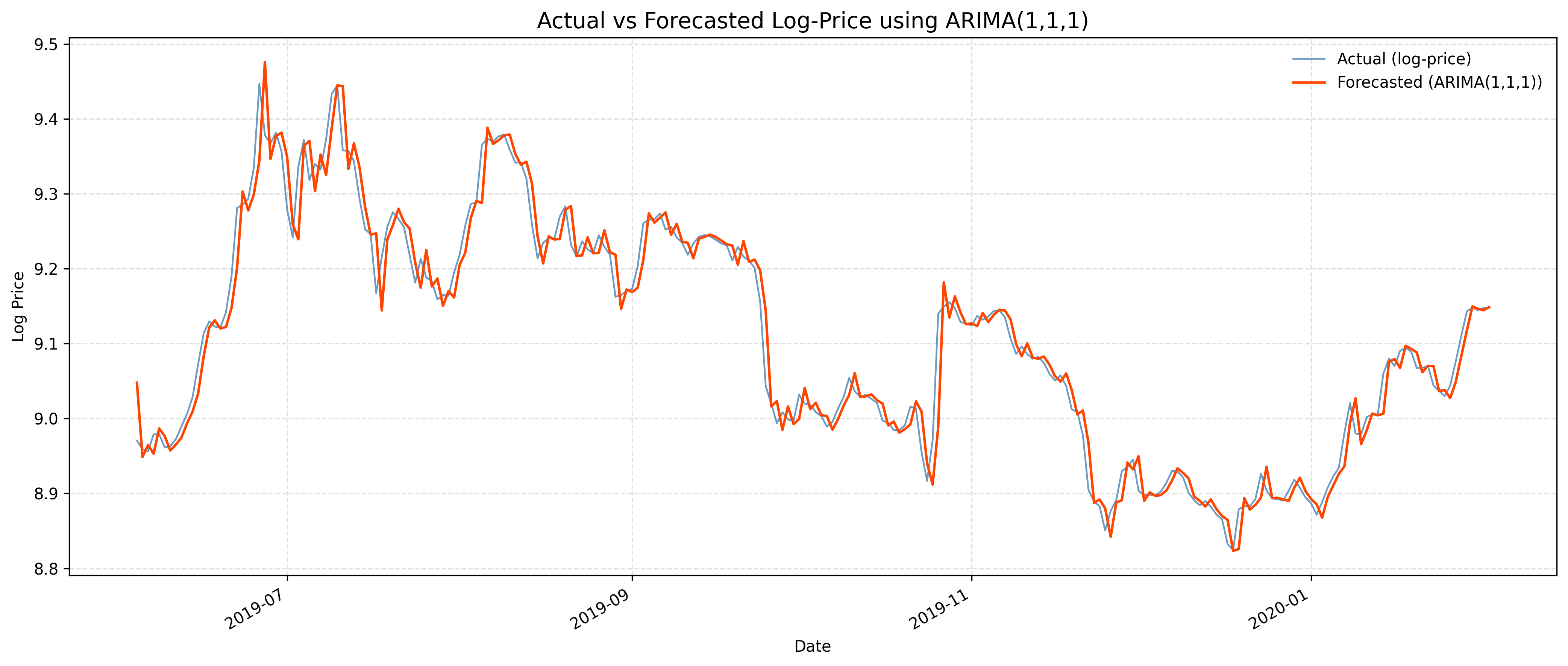}
    \caption{Actual vs Forecasted Log-Price using ARIMA(1,1,1).}
    \label{fig:forecasting}
\end{figure}

As shown in Figure~\ref{fig:forecasting}, the ARIMA(1,1,1) model captures the dynamics of the Bitcoin log-price reasonably well. 
The forecasted values (red line) track the actual log-prices (blue line) closely, especially in capturing both upward and downward trends. 
While the model smooths out sudden jumps and extreme volatility, a common limitation of ARIMA models, it is effective in modeling the gradual changes and short-term dependencies of the data. 
Overall, the forecast demonstrates that ARIMA(1,1,1) provides a strong approximation of Bitcoin's log-price movements, offering useful insights into its short-term behavior.

\section{Conclusion}
This study evaluated the performance of classical time series models in forecasting Bitcoin prices by applying ARIMA, SARIMA, GARCH, and EGARCH to daily data. 
Each model was trained on the first 90\% of the dataset and tested on the final 10\%, allowing for a clear assessment of out of sample predictive accuracy. 
The results showed that ARIMA(1,1,1) provided the strongest forecasts for short term log price dynamics, outperforming SARIMA in terms of MAE and RMSE, while SARIMA’s seasonal terms did not deliver notable improvements. 
On the volatility side, the EGARCH(1,1) model achieved the best fit, as reflected in lower information criteria values compared to the basic GARCH model, and successfully captured the asymmetric response of volatility to negative and positive shocks. 
These findings highlight that even simple and well established statistical models can capture meaningful features of Bitcoin’s behavior, including both price dynamics and volatility clustering.

At the same time, the analysis revealed the limitations of these methods. 
ARIMA forecasts tended to smooth extreme price jumps, and GARCH models, while effective at modeling volatility persistence, were less able to fully reflect sudden spikes in risk. 
This underlines the challenge of forecasting an asset as volatile and rapidly evolving as Bitcoin. 
Nevertheless, the study demonstrates that classical models remain useful for short run forecasting and provide a strong foundation for further research. 
Future studies may build on these results by incorporating machine learning techniques or by extending the analysis to include macroeconomic and blockchain variables. 
While no model can perfectly predict Bitcoin, continued investigation will help move closer to more reliable forecasting strategies for highly volatile financial assets.

\newpage
\section*{References}

\noindent\textbf{\label{ref:pichl2017}Pichl, L., \& Kaizoji, T. (2017).} 
Volatility analysis of Bitcoin price time series. 
\textit{Quantitative Finance and Economics, 1}(4), 474–485. 
\href{https://doi.org/10.3934/QFE.2017.4.474}{https://doi.org/10.3934/QFE.2017.4.474}

\vspace{0.5cm}

\noindent\textbf{\label{ref:yenidogan2018}Yenidoğan, İ., Çayir, A., Kozan, O., Dağ, T., \& Arslan, Ç. (2018, September).} 
Bitcoin forecasting using ARIMA and PROPHET. 
In \textit{2018 3rd International Conference on Computer Science and Engineering (UBMK)} (pp. 621–624). IEEE. 
\href{https://doi.org/10.1109/UBMK.2018.8566413}{https://doi.org/10.1109/UBMK.2018.8566413}

\vspace{0.5cm}

\noindent\textbf{\label{ref:mudassir2020}Mudassir, M., Bennbaia, S., Unal, D., \& Hammoudeh, M. (2020).} 
Time-series forecasting of Bitcoin prices using high-dimensional features: A machine learning approach. 
\textit{Neural Computing and Applications, 33}(15), 9155--9177. 
\href{https://doi.org/10.1007/s00521-020-05129-6}{https://doi.org/10.1007/s00521-020-05129-6}

\end{document}